# Topological scattering singularities and embedded eigenstates for polarization control and sensing applications

Zarko Sakotic[1*], Alex Krasnok[2], Andrea Alú[2,3], and Nikolina Jankovic[1]

[1]BioSense Institute-Research Institute for Information Technologies in Biosystems, University of Novi Sad, Dr Zorana Djindjica 1a, 21101, Novi Sad, Serbia

[2]Advanced Science Research Center, City University of New York, New York, NY 10031, USA

[3]Physics Program, Graduate Center, City University of New York, New York, NY 10016, USA

Epsilon-near-zero and epsilon near-pole materials enable reflective systems supporting a class of symmetry-protected and accidental embedded eigenstates (EE) characterized by a diverging phase-resonance. Here we show that pairs of topologically protected scattering singularities necessarily emerge from EEs when a non-Hermitian parameter is introduced, lifting the degeneracy between oppositely charged singularities. The underlying topological charges are characterized by an integer winding number and appear as phase vortices of the complex reflection coefficient. By creating and annihilating them, we show that these singularities obey charge conservation, and provide versatile control of amplitude, phase and polarization in reflection, with potential applications for polarization control and sensing.

## 1. Introduction

Along with the progress in fabrication technologies observed in the last few decades, the concepts of metamaterials [1], metasurfaces [2] and photonic crystals [3] have brought forward a previously unattainable level of control of visible, infrared and microwave electromagnetic waves. More recently, the pursuit of wave control has been expanded with tools borrowed from the field of topology [4], enabling phenomena such as robustness to imperfections and immunity to back-scattering [5,6], as well as unidirectional transport [7]. Topological photonics has been rapidly growing in recent years, transferring established concepts from condensed-matter systems to electromagnetics research and metamaterials. Topological systems are usually quantified with an invariant – an integer-valued quantity, like the Chern number [8], which does not change upon continuous deformations that preserve the topological nature [9]. Although topological aspects of photonic systems have mostly been driven through the condensed matter physics perspective, recently there has been an expansion of topological photonics beyond these boundaries, leveraging the distinct features of photons [10,11]. Contrary to the fermionic nature of electrons, photons have a bosonic nature, thus offering different opportunities for the realization of topological phenomena [12]. Specifically, topological effects are increasingly being found in scattering and radiative processes [13], moving beyond the limitations of the tight-binding model that commonly models electronic systems [11]. Bound states in the continuum (BIC) or embedded eigenstates (EE), examples of peculiar features associated with electromagnetic radiation and scattering, provide topological features to the optical response rooted in modes that are non-radiative yet embedded within the radiation continuum [14-21].

Photonic EEs supported by periodic systems have indeed been shown to possess topological features in the form of a polarization singularity in the wavevector space. Their robustness has been explicitly attributed to their topological nature [22,23], rooted in the fact that these singularities comply with topological charge conservation [24,25]. The topological nature of EEs has spun-off several research efforts, exploring the merging of EE charges to produce even more confined resonances in realistic systems [26] and unidirectional guided modes within the continuum [27]. The topological properties of EEs have been especially useful for polarization control, as it was shown that topologically protected polarization conversion is possible [28-30], and circularly polarized states can arise from BICs by breaking spatial symmetries [31-33]. Recently, the generation of vortex beams through EEs [34] and efficient topological vortex laser generation [35] were demonstrated, showing the potential of topological phenomena in radiative and scattering processes. Further connection between novel topological phenomena in the form of higher-order corner states and EEs has also recently been established in [11], indicating the far-reaching topological consequences that non-radiating states may have on the topological states of electromagnetic structures. Besides using the standard Hamiltonian formalism, the topological features of these systems can be studied through the scattering matrix formalism, and specifically by analyzing the complex reflection coefficient, where phase vortices arise [28, 36-37]. This approach is very valuable because the features of the scattering matrix correspond to actual observables that can be looked for experimentally.

Although the topological aspects of EE-related phenomena are well understood in periodic systems, there has been little exploration into the topological features of other photonic systems that support EEs. Specifically, EEs arising in structures with singular values of the permittivity, mainly using epsilon-near-zero (ENZ) materials, have been recently studied [38-41], showing that they enable versatile optical and thermal emission properties [42, 43]. However, their topological nature has not been discussed yet, which may further boost their potential in photonic and thermal applications. Furthermore, scattering anomalies such as EEs can be well studied in the complex frequency plane [44], which has not been applied so far to the analysis of topological scattering phenomena.

In this paper, we extend the concept of topological photonics to planar reflective systems that support EEs and, using the scattering matrix formalism and complex frequency analysis, we unveil their topological nature and the emergence of topologically-protected scattering singularities. The proposed system supports symmetry-protected and accidental EEs, which are shown to be the origin of scattering singularities emerging upon insertion of loss/gain in the constituent materials. We focus on the lossy case and show that perfect-absorption singularities are intrinsically connected to the underlying EEs. Methods of formation, annihilation and control of these topological charges are discussed, providing a versatile tool to manipulate amplitude, phase and polarization of reflected waves. Using these concepts, we demonstrate several applications of these phenomena for polarization control and sensing.

## 2. Results and discussion

### A. Symmetry protected embedded eigenstates and emerging singularities

We consider a basic planar structure, infinitely extended in two dimensions as shown in the inset of Figure 1(a) and analyze the leaky mode dispersion of a PEC-backed slab with Drude permittivity dispersion $\varepsilon_1 = \varepsilon_0(1 - \omega_p^2/(\omega^2 + j\gamma\omega))$ in the proximity of the plasma frequency $\omega_p$, Figure 1(a). Transverse magnetic (TM) bulk modes in a plasma slab have been well-studied in the literature [45]. However, only recently it has been realized that such a slab supports a symmetry-protected EE in the lossless limit ($\gamma = 0$) [43,46].

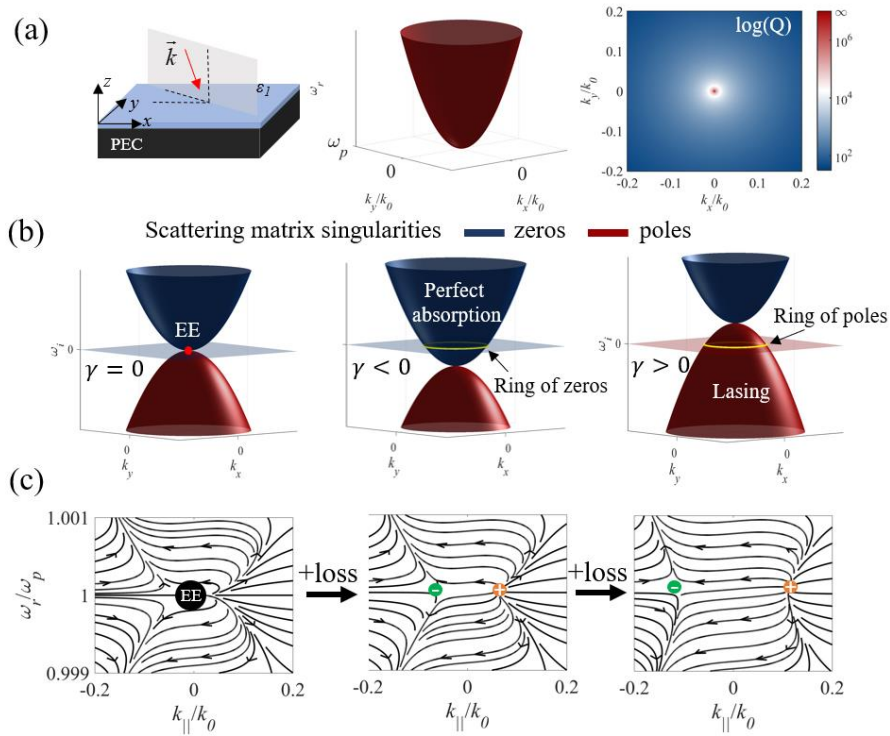

**Figure 1.** (a) Sketch of a PEC-backed slab with Drude dispersion; Bulk mode dispersion for TM polarized light. At $k_x = k_y = 0$ and $\omega_p$ the system supports a symmetry-protected EE, as indicated by the Q-factor divergence on the right. (b) Dispersion of scattering matrix singularities in the imaginary frequency space – zeros (perfectly absorbing states) and poles (eigenmodes); Lossy or gainy Drude model enables the creation of a ring of perfect absorption/lasing points at the real frequency axis. (c) Vector flow of the complex reflection coefficient $r_{TM}$ in the wave number-frequency plane. The EE splits into two vortices – a saddle point and a source point - which move away from each other as loss increases.

Namely, for zero transverse wavenumber $k_\parallel = 0$ at frequency $\omega_p$, the quality factor of such a mode diverges, Figure 1(a) (right panel), and the charges in this self-sustained mode of the slab oscillate along the $z$-direction in the entire volume ($E_{x,y}=0$, $E_z \neq 0$). Consequently, an incoming wave impinging at normal incidence ($E_z^{inc}=0$) cannot excite this dark mode, in accordance with reciprocity, since the mode is not coupled to the radiation continuum.

In order to obtain a better picture of the leaky nature of these modes in this basic geometry, we analyze the problem in the complex frequency space $\omega=\omega_r+i\omega_i$, which enables a detailed and intuitive description of the underlying scattering phenomena [44]. We consider the $e^{-i\omega t}$ time convention and compute the scattering matrix eigenvalues in the complex frequency plane. The singularities of the scattering matrix eigenvalues, i.e., poles and zeros, have a straightforward interpretation – poles correspond to source-free solutions of Maxwell's equations and correspond to the system eigenmodes as purely outgoing waves, while zeros of the scattering matrix eigenvalues correspond to modes with purely incoming nature [47]. These singularities are connected via complex conjugation in Hermitian (lossless) systems, and correspond to the time-reversed version of each other [48].

Since the structure under consideration is backed by a PEC, it can be described by a one-port network, whose scattering matrix has only one element – the reflection coefficient, $S = [r]$. Thus, computing the reflection singularities is sufficient in this scenario, which can be achieved using a transmission line model for the structure (Supplementary Materials), from which the conditions for zero/pole are derived as

$$\tanh(jk_{1z}t) = \pm \frac{Z_0}{Z_1}, \qquad (1)$$

where $t$ is the layer thickness, $k_{1z} = \sqrt{k_0^2 \varepsilon_1 - k_x^2 - k_y^2}$ is the wavenumber along the $z$-axis in the Drude material, and $Z_0 = k_{0z}/\omega\varepsilon_0$ and $Z_1 = k_{1z}/\omega\varepsilon_0\varepsilon_1$ are the TM wave impedances in air and the Drude material, respectively. The plus sign refers to the reflection-zero condition, while the minus sign is the reflection-pole condition. More details on the derivation are given in the Supplementary Material.

Figure 1(b) shows the dispersion of the poles and zeros (red and blue) of $r$ in the $\omega_i - k$ parameter space. In the lossless case, the parabolic dispersion of poles and zeros mirror each other due to Hermiticity ($\omega_i^{pole} = -\omega_i^{zero}$), and they touch on the real frequency axis ($\omega_i = 0$) for zero transverse wavenumber $k_{//}=0$, forming an EE. By introducing loss (gain) in the slab, the dispersion surfaces translate down (up) along the imaginary frequency axis, creating a ring of real-frequency zeros (poles). The intersection of scattering singularities with the real frequency axis represents a topological object – a charge with non-zero winding number of a physical parameter in given parameter space. The topological signature of these states can be observed in the wavenumber-frequency plane by plotting the vector flow of $r$, Figure 1(c). In the $\omega_r - k_\parallel$ two-dimensional space, taking into account a single incidence plane, it is easily observable that, upon adding loss, the EE splits into two vortices, a saddle-type (green) and a source type (orange). We only use loss in the following discussion to control the non-Hermiticity of the system, since loss is unavoidable while gain cannot be as readily obtained in most practical scenarios, but similar considerations may be applied to gain as well. As we will discuss later, the non-zero winding number associated with these vortices arises in the phase of the reflection coefficient in the real-frequency space, as phase acquires a $\pm 2\pi$ increment when encircling them in the frequency-incident angle space.

These types of vortices have been studied extensively in the context of singular optics [49,50]. The splitting of EE into a pair of charges, in this case a pair of perfectly absorbing points, sheds a new outlook to previous reports on *pairs* of perfect absorption occurring in reflective systems [36,51-52]. As we show later, each one of the analyzed EEs produces exactly two charges on the $k_\parallel$ axis (incidence angle $\theta = \sin^{-1}|k_\parallel/k_0|$), due to the parabolic dispersion of the scattering singularity around the EE. These perfectly absorbing states can only be destroyed by charges of opposite polarity if they are available. In the case studied in Figure 1(c), there are no additional charges available, thus there is no mechanism to annihilate charges except going back to an EE in the lossless limit ($\gamma = 0$), which represents a special case of merging charges. This merging is, in fact, necessary due to energy conservation, as no energy can be absorbed in a lossless system. Another way to interpret this result is to notice that perfect absorption (reflection-zero) corresponds to the condition of *critical coupling*, where absorption and radiation losses are perfectly matched. If absorption losses $\gamma$ turns to zero, radiation losses must also turn to zero to yield a singularity, and the eigenstate is *decoupled* from the environment, thus showing the close correlation between critical coupling and EEs in one port systems.

## B. Accidental EEs, creation and annihilation of topological charges, and their phase signatures

To further enrich the discussion, we move to a more general scenario and analyze the system shown in Figure 2(a), consisting of a dielectric spacer sandwiched between a PEC and a resonant top layer. We model the top layer with a Lorentz type of permittivity response, as shown in Figure 2(b):

$$\varepsilon_1 = \varepsilon_0 \left(1 + \frac{\omega_p^2}{\omega_0^2 - \omega^2 - j\gamma\omega}\right) \qquad (2)$$

where $\varepsilon_0$ is the vacuum permittivity, $\omega_p$ is the plasma frequency, $\omega_0$ is the Lorentz resonance frequency, and $\gamma$ is the damping or absorption loss. This type of permittivity describes the electric response of different naturally occurring materials, including metals, semiconductors and polar dielectrics like SiC or AlN [53]. Moreover, such a response can be induced by resonant metasurfaces, either with metallic or dielectric realizations [36, 54]. Thus, our theoretical discussion can be applied to various scenarios, whether using isotropic bulk materials, 2D materials or metasurfaces.

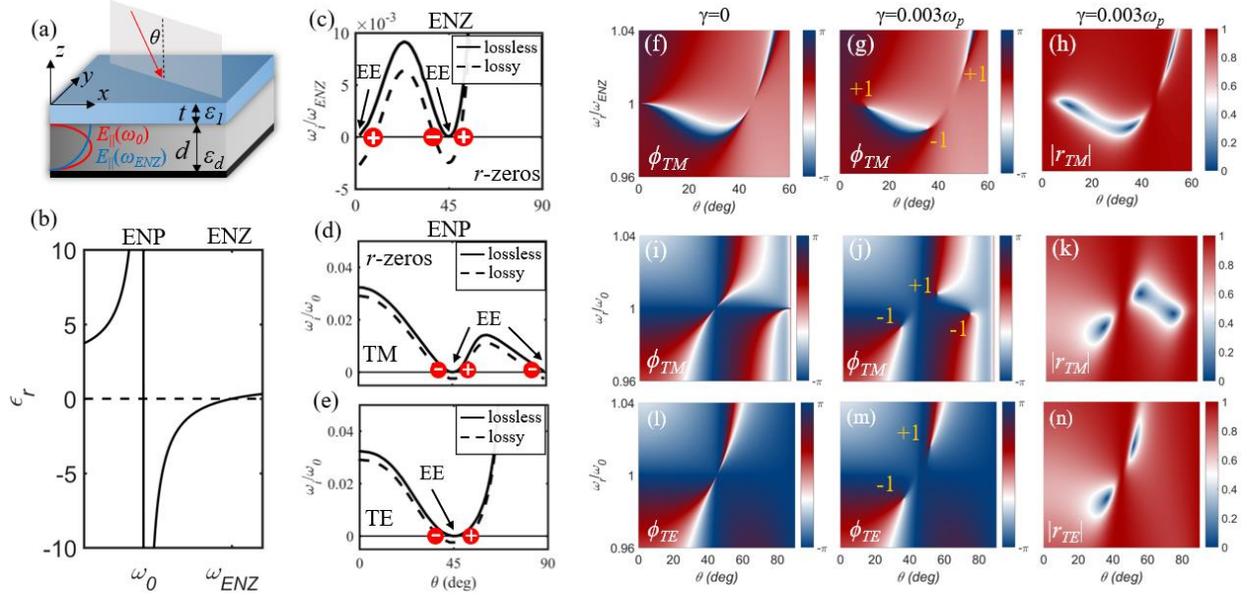

Figure 2. (a) Sketch of the planar multilayer structure under oblique illumination. (b) Permittivity dispersion of the top layer. (c)-(d) Reflection-zero dispersion in the ENZ and ENP regions. For the ENZ case, thicknesses were chosen as $t/\lambda_{ENZ}=0.05$ and $d=\lambda_{ENZ}/4$, and for the ENP case $t/\lambda_{ENZ}=0.007$ and $d=\lambda_0/2$. Dielectric permittivity is $\varepsilon_d=1$. (f-h) Phase and amplitude plots of TM reflection in lossless and lossy case for ENZ EEs. (i-k) Phase and amplitude of TM reflection in lossless and lossy case for ENP EEs. (l-n) Phase and amplitude of TE reflection in lossless and lossy case for ENP EEs.

The spectral points of particular interest are the Lorentz resonance frequency $\omega_0$ and zero-crossing $\omega_{ENZ}$ frequency. In the lossless case ($\gamma = 0$), the permittivity attains singular values at these frequencies, $|\varepsilon|=\infty$ and $\varepsilon=0$, respectively. We refer to these regions as ENP (epsilon near-pole) and ENZ [55]. Under these conditions, such a layer imposes a hard boundary condition [17] – at $\omega_0$ it effectively acts like an electric wall or PEC (for both transverse-electric TE and transverse-magnetic TM polarizations), while at $\omega_{ENZ}$ it acts like a magnetic wall or PMC (for TM polarization only). This can be attributed to the surface impedance of the layer going to 0 (PEC at $\omega_0$) or infinity (PMC at $\omega_{ENZ}$). With the bottom PEC, such a theoretical structure can provide perfect light trapping at the frequencies of singular permittivity, since the top and bottom layers act like perfect mirrors.

The dielectric spacer supports a continuum of modes above the lightline, which can couple to free-space radiation, as this structure is electromagnetically open from the top. However, if a mode of the spacer, i.e., a Fabry-Perot mode, overlaps with one of the top layer material resonances, a leaky mode with zero radiation decay (EE) is supported [38]. At $\omega_0=2\pi f_0$, the spacer thickness required to support such a mode is

$$d = \frac{n\lambda_r}{2} = \frac{nc}{2f_0\sqrt{\varepsilon_d - \sin^2\theta}}, \qquad (3)$$
$$n = 0,1,2\ldots$$

where $c$ is the speed of the light in vacuum, $f_0$ is the Lorentz resonance frequency, $\varepsilon_d$ is the dielectric layer permittivity, and $\theta$ is the incidence angle. The tangential component of the electric field is required to be zero at both top and bottom boundaries, as sketched in the inset of Figure 2(a). On the other hand, at $\omega_{ENZ}=2\pi f_{ENZ}$ the magnetic field has a null at the boundary with the top layer, thus the resonant thickness is

$$d = \frac{(2n+1)\lambda_r}{4} = \frac{(2n+1)c}{4f_{ENZ}\sqrt{\varepsilon_d - \sin^2\theta}}, \qquad (4)$$
$$n = 0,1,2\ldots$$

We first focus on the ENZ case – Figure 2(c) shows the dispersion of the reflection-zeros in $\omega_i$-$\theta$ plane. We note here that the reflection-zero condition is identical to the condition of perfect absorption, perfect impedance matching, and critical coupling, as all of these describe the same phenomenon in one-port systems. Like in the previous discussion, the reflection coefficient was obtained using the transmission line method (Supplementary Materials). At the plasma frequency, this structure supports a symmetry-protected EE at 0 deg and an accidental EE at a designed angle of 45 deg. Namely, the dispersion of the resonant mode is "pinned" to the point of normal-incidence and ENZ frequency for any dielectric thickness, thus this symmetry-protected EE is intrinsic to this configuration and cannot be controlled or moved. On the other hand, the accidental EE can freely move along the incident angle axis at $\omega_p$, and the angle of EE can be chosen according to Eq. (4).

Although the magnitude of the reflection coefficient is unity for any frequency and angle in the lossless case, the EEs are visible in the reflection coefficient phase. In real-frequency space, they manifest themselves as phase resonances with diverging linewidth, as visible in Figure 2(f). These features are only available for TM excitation due to the plasmonic nature of the underlying modes. A similar result with diverging phase-resonances was reported in [56], although the origin of EE here is quite different.

The utility of phase analysis becomes apparent when losses are introduced, Figure 2(c, f-h). Namely, the dispersion of reflection-zeros shown in Figure 2(c) shifts down along the imaginary frequency axis, creating intersections with the $\theta$-axis and thereby creating topological charges. These charges appear as phase vortices, whose charge and polarity is defined by the amount of phase accumulation when encircling them counter-clockwise in the plane:

$$q = \frac{1}{2\pi}\oint d\phi. \qquad (5)$$

As shown in the amplitude plot, pairs of reflection-zeros and consequently perfect absorption points emerge from EE, where the charge emerging from the symmetry-protected EE has a mirror copy in the negative $\theta$-axis and thus is not visible in the plot.

A similar scenario arises for the ENP case – an accidental EE appears at a designed angle of 45 deg. More interestingly, however, both TM and TE modes have real-eigenfrequencies at the same point in the dispersion diagram, Figure 2(d) and (e). This happens because at $\omega_0$ the impedance of the top layer goes to zero, essentially acting as a perfect mirror regardless of polarization. Additionally, for TM polarization a symmetry-protected EE appears at the light line (90 deg incidence) and, analogously to the ENZ case, it is "pinned" to the same point regardless of the spacer thickness. We again introduce loss and notice topological charges emerging. Symmetry protected EE produces one charge with the oppositely charged mirror copy in the negative $\theta$ space (due to symmetry), while the accidental EEs split into two charges. It should be noted that the ENP layer used as a mirror here has a finite thickness, and thus hosts an infinite number of FP modes in the lossless case as the permittivity approaches $+\infty$, for $\omega \to \omega_0^-$. However, for structures and materials considered here, these modes are suppressed, and their effect on the scattering properties can be disregarded as the response is dominated by the spacer mode (see Supplementary Materials). Another remarkable difference between the ENP and ENZ cases is that the former can produce perfect absorption points at normal incidence, whereas ENZ requires a non-zero incident angle to engage plasmons and absorb waves. More details on the scattering properties in the ENP case can be found in Supplementary Materials.

To verify the conservation of topological charge, we introduce a mechanism to annihilate charges. As evident from the reflection-zero dispersion diagrams in Figure 2(c)-(e), non-zero loss $\gamma$ creates and pushes closer together charges originating from neighboring EEs. It is then reasonable to assume that a further increase of loss can eventually bring these charges together, causing their mutual annihilation. This would be equivalent to moving the reflection-zero dispersion in the lower complex half-plane, thus removing the intersection with the $\theta$-axis, Figure 2(c)-(e). However, instead of changing the absorption loss in the material, we can induce more radiation loss by lowering the Q-factor of the underlying eigenmode. This can be done by reducing the top layer thickness, which essentially reduces its reflectivity (weaker "light-trapping") and consequently moves charges closer to each other until they eventually annihilate.

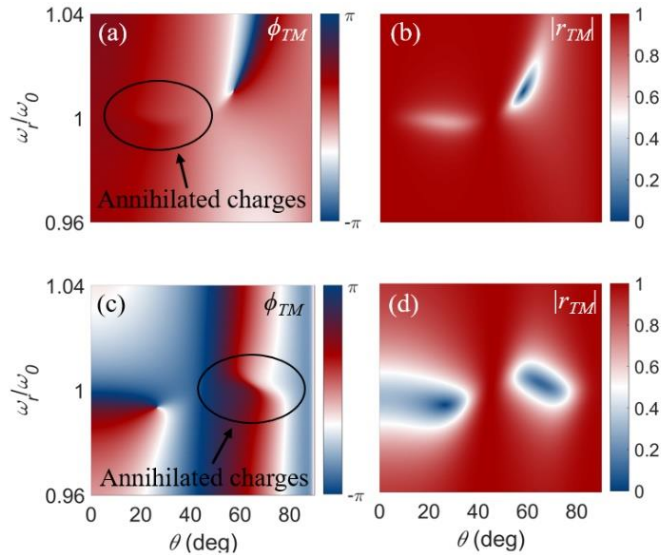

Figure 3. (a) Phase of reflection for the ENZ case. Two charges annihilated and one remaining. (b) The reflection zeros associated with these charges have vanished. (c) Same as (a) for ENP case. (d) Same as (b) for ENP case.

Figure 3 shows the ENZ and ENP cases for TM polarized light with reduced top-layer thickness. Comparing these results to Figure 2(g-h) and (j-k), the neighboring charges have been annihilated and consequently the reflection-zeros associated with them have vanished. A useful rule-of-thumb can be inferred here – the angles at which EEs arise represent vertical walls impenetrable to these charges, and charges can move and annihilate only with the ones originating from neighboring EEs. For example, the remaining charge in the ENZ case in Figure 3(a) and (b) cannot be destroyed for any level of material loss and thickness. Although there is a mirror copy of this charge in the negative $\theta$ half-plane due to the symmetry, these charges cannot merge since they are separated by "impenetrable" EE-walls. The only way to annihilate this charge is to increase the resonator length such that second-order Fabry-Perot mode EE appears, and its associated charge is brought into contact with the other charge (supplementary materials). In the ENP case in Figure 3(c) and (d), however, the remaining charge and its mirror copy in the $\theta$-plane do not have an EE between them, and they are free to annihilate for a proper amount of radiation (or absorption) loss.

This result opens the question of material loss – indeed, some of the naturally occurring materials have high absorption loss and consequently some features discussed here are not available, i.e., charges may already be destroyed due to large loss. However, if we constrain the discussion to low-loss or moderately lossy materials, e.g., polar dielectrics like SiC, all of the previously described aspects hold. Furthermore, artificial materials as well as 2D materials, can provide ENZ or ENP response, thus expanding the validity of the presented analysis to a wide range of realistic geometries. For example, a Lorentzian (ENP) response can be induced as an electric dipole resonance in metasurfaces [36,51] or graphene strips [57], where similar features have been observed.

It is worth noting that the accidental EE at the designed angle $\theta$, and its mirror copy at $-\theta$, are part of a *ring* of EEs in *k*-space, as they represent intersection points of a chosen incidence plane and the EE-condition in the whole $k_\parallel$ plane. This condition, which can be derived from Eqs. (3) and (4), is described by a circle, whose radius is defined by the spacer thickness:

$$\left(\frac{k_x}{k_0}\right)^2 + \left(\frac{k_y}{k_0}\right)^2 = \sin^2\theta = \begin{cases} \varepsilon_d - \left(\frac{n\lambda_0}{2d}\right)^2, k_0 = \frac{2\pi}{\lambda_0} \text{ for the ENP case.} \\ \varepsilon_d - \left(\frac{n\lambda_{ENZ}}{4d}\right)^2, k_0 = \frac{2\pi}{\lambda_{ENZ}} \text{ for the ENZ case.} \end{cases} \quad (6)$$

This is in contrast to EEs in photonic crystals, where accidental EEs arise at isolated *k*-points. This property implies that EEs here do not represent singularities in *k*-space, and thus have different features than EEs in periodic systems.

The discussion here adds new insights on previous observations of phase singularities and perfect absorption in metasurfaces [36], establishing a connection between topologically-protected scattering features and EEs, and showing that properly designed planar structures can host a variety of topological scattering phenomena. The interplay of material and radiation losses provides control of the singularities and enables a versatile platform for the control of intensity and phase in reflection, which is important in applications like thermal engineering, polarization control and sensing.

## C. Topological charges near EE for polarization control

So far, our discussion has been constrained to the analysis of phase and amplitude of the reflection coefficient around EEs. However, an important aspect intrinsically connected to these features is polarization. Traditionally, manipulation of the polarization state of light has been based on waveplates [58]. When light propagates through birefringent crystals, orthogonal linear polarizations (LP) experience different absorption coefficients and phase accumulation, enabling the generation of purely horizontal, vertical or circular polarization (CP) at the output.

As we have shown in the previous discussion, the system of Figure 2 displays co-located EEs with TE and TM polarization in the ENP case, and after accounting for loss, reflection-zeros of both polarizations emerge in the vicinity of each other. This property has interesting consequences for polarization control, as the two orthogonal polarization states experience dramatically different absorption coefficients and phase accumulation in the region containing these charges, allowing for different linear transformations of the polarization state upon reflection. To test the potential of this property, we use isotropic SiC to model the top layer permittivity [59]. We choose *t*=300 nm for the SiC layer, and the dielectric layer thickness is calculated according to Eq. (2) at the transverse optical phonon frequency $\omega_{TO} = 2\pi$ 23.89 THz for the angle of 50 deg. Namely, the ENZ and ENP regions are available at the longitudinal and transverse optical phonon frequencies of SiC. We plot the ellipsometric parameter $\tan^{-1}(r_{TM}/r_{TE})$ to visualize both charges, Figure 4(a), where the maximum value in the density plot represents the TE-zero, while the minimum represents the TM-zero. It is clear that these two points in the parameter space can work as polarization filters or polarizers – for a mixed polarization input, only TM or TE polarized light exits the structure. This is in contrast to the ENZ case, where only the TM-zero is available - we note that this was the basis for the polarization switching scheme studied in [60]. Beyond admitting zeros of both polarizations, the phase difference between orthogonal polarizations induced by the structure also dramatically changes around the charges, as they represent phase vortices. This property opens up possibilities for generation of various polarization states in the output for a mixed input – for example, converting LP to CP light. To analyze these possibilities, we apply standard Jones calculus and express the input and output Jones vectors as

$$J^i = \begin{bmatrix} a \\ be^{j\delta_i} \end{bmatrix}, \quad J^o = \begin{bmatrix} r_{TE}a \\ r_{TM}be^{j\delta_i} \end{bmatrix} = \begin{bmatrix} A \\ Be^{j\delta_o} \end{bmatrix} \quad (7)$$

$$\delta_r = \delta_o - \delta_i \quad (8)$$

where $a$, $b$, and $\delta_i$ are the input amplitudes and phase difference of orthogonal polarizations, $A$, $B$, and $\delta_o$ being the output amplitudes and phase difference of orthogonal polarizations, and $\delta_r$ represents the phase retardation between orthogonal polarizations that the structure introduces. The reflection matrix used here has zero off-diagonal elements since there is no polarization cross-coupling mechanism.

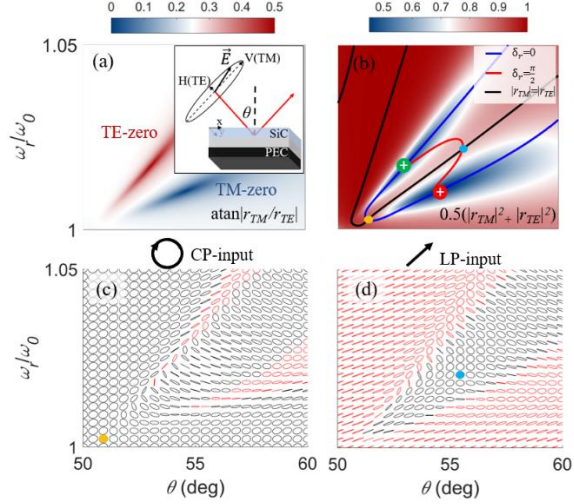

Figure 4. Polarization manipulation. (a) Ellipsometric parameter $\psi = \tan^{-1}(r_{TM}/r_{TE})$ normalized to $\pi$. (b) Total reflectance as an efficiency indicator. (c) Polarization ellipses in the output for CP-input. Red and black colors indicate handedness (red for RCP, black for LCP). (d) Polarization ellipses in the output for 45 deg LP-input.

To illustrate the opportunities for polarization control, we plot the total reflectance and the contour lines with specific $\delta_r$ values. In this way, we can find conversion points from LP to CP and vice versa, Figure 4(b). Namely for $a=b$, and $|r_{TE}| = |r_{TM}|$ the output amplitudes are equal, $A=B$. Thus by finding the intersection of $|r_{TE}| = |r_{TM}|$ and $\delta_r = \pi/2$, we have a LP to CP conversion point and vice versa. This is represented by the blue dot in the figure. The orange dot represents the point that preserves polarization, as $|r_{TE}| = |r_{TM}|$ and $\delta_r = 0$, and the green and red circles represent phase vortices. Thus, the proposed system has polarization conservation, filtering, and LP↔CP conversion capabilities, all in the EE proximity.

In order to better visualize this effect, we plot the output polarization ellipses for both CP and 45 deg LP inputs, Figure 4(c) and (d). Singular phase points produce purely vertical or horizontal polarization states in the output. This can be done efficiently since most of the desired polarization is reflected, while the other is fully absorbed. However, extracting horizontal or vertical polarization from a 45 deg LP or CP input is limited to 50% in efficiency, as no conversion from one to the other happens, $r_{TE-TM} = r_{TM-TE} = 0$. CP to LP and LP to CP conversion is also possible (blue dot), preserving around 80% of incident power. Due to the vicinity of the different polarization features, the output polarization state is sensitive to small changes in the system, and specifically to the permittivity (loss) of the SiC layer (Supplementary Materials). This could lead to a versatile polarization switching platform based on active control of material permittivity - for example, a moderate change in permittivity of InAs in the Reststrahlen region was recently demonstrated using laser-induced nonlinear processes [61]. Polarization control in the long-wave infrared range is especially important, as birefringent materials are scarce in this range of the electromagnetic spectrum [62], and the presented topological features using SiC and other IR materials may help circumvent this challenge. More details about the dependence of output polarization state on material losses can be found in Supplementary Materials.

### D. Singular phase and near-annihilation point for sensing applications

Phase vortices are characterized by an undefined phase point in their center, around which the phase changes dramatically. This feature, most commonly observable around reflection-zeros, has been used as a basis for interferometric phase sensing schemes [63-68]. Namely, even small changes in the environment can result in giant phase changes, thus creating one of the most sensitive schemes. Various systems exhibit these types of singular points, including metasurfaces [63], 1-D photonic crystals [64], hyperbolic [65] and 2D materials [66]. Instead of using a traditional interferometric setup, most of these schemes use ellipsometric measurements, which measure phase and amplitude differences between orthogonal linear polarization components.

The caveat of this sensing method is the following: by approaching the singular point of the vortex, phase changes more dramatically, thus increasing the sensitivity of the system. However, the amplitude of the reflection coefficient drops at the same time. The sensitivity diverges as the singular point is reached, but there is no reflection to be measured at the detector. This inverse relationship between sensitivity and |r| plays an essential role in such devices.

There are indeed qualitative differences between different realizations. Engineering structures with lower possible values of |r| lead to higher sensitivity. Furthermore, higher Q-factors of the underlying mode can improve the sensitivity, as these modes produce stronger fields and interactions with the environment. A critical issue to reconcile in such devices is the need for a highly precise angle of incidence to engage abrupt phase jumps – e.g., extremely high sensitivity was reported in [66]. However, the angle required for such performance requires precision on the order of $10^{-4}$ degrees, and any deviation from the exact angle reduces the sensitivity. Thus, the sensitivity of such devices is hard to predict and maintain stable.

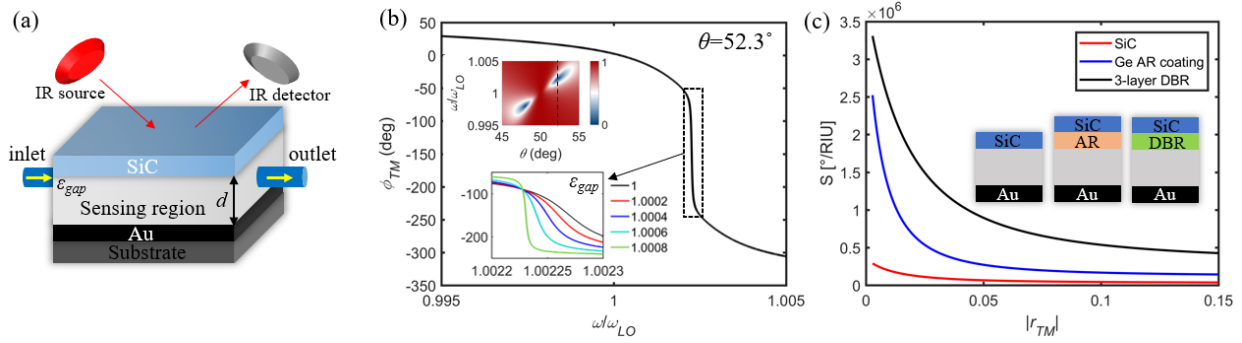

Figure 5. (a) Sensing scheme based on phase singularities. (b) Phase of the reflection coefficient as the frequency is changed and passes near the vortex. Inset: $|r_{TM}|$ near the EE at 50 deg; Phase sensitivity to small changes of the gap permittivity. (c) Sensitivity for three different configurations – improvement due to enhanced Q-factor with antireflection coating and DBR.

To address these issues, we exploit the unusual physics around the EE and the related topological charges discussed in this paper. First, we show that using EEs in SiC provides high-Q factors and enables remarkably high sensitivities around the reflection-zeros associated with EEs. Secondly, we introduce the concept of *near-annihilation point* sensing and show that we can alleviate the requirement for an extremely precise angle of incidence by manipulating charges, thus creating an extended angular range with stable sensitivity.

To achieve these goals, we envision a gold-backed air gap (spacer) covered by a SiC layer to be used as a trace gas sensor, operating at long-wave infrared wavelengths, Figure 5(a). Gold is highly reflective at longwave IR wavelengths, thus it can work as an efficient reflective bottom layer. For the chosen gap size $d$=32 μm, several higher-order FP modes are engaged. However, the spectrum is not overcrowded, due to the long operating wavelength of 10.3 μm. Furthermore, having a gap this size is advantageous because a nanofluidic channel can also be integrated to function as a spacer, therefore expanding the possible applications of the proposed structure.

As shown in the previous discussion, the SiC-capped resonator can support an EE around 10.3 μm (ENZ region) and the accompanying reflection-zeros. To demonstrate the phase jump, we plot the reflection coefficient near EE and its phase as it passes near the vortex at 52.3 deg, Figure 5(b). Based on this phase jump, we can detect tiny changes in the sensing layer – as indicated in the bottom left inset in Figure 5(b). At a constant frequency and angle, the phase of reflected waves changes dramatically for permittivity changes on the order of $10^{-4}$, producing very high sensitivity, of the order of $10^5$ deg/RIU. However, as mentioned before, the sensitivity is highly dependent on the amplitude of the reflection coefficient, i.e., the proximity of the measurement to the singular point. To better grasp this correlation, we plot the sensitivity calculated at different values of |r|, red curve in Figure 5(c). As expected, we see an increase in sensitivity as /r/ approaches zero. However, instead of relying on extremely small values of |r| to get higher sensitivity, which could make the signal at the detector indistinguishable from noise, we can improve the sensitivity by increasing the $Q$-factor of the mode instead. This can be done by placing a high-index anti-reflection coating or a distributed Bragg reflector (DBR) between the spacer and SiC, improving the reflectivity of the top layer. To this end, we can use low-loss longwave IR materials, for example, Ge and $BaF_2$, as high and low index materials, respectively. As shown in Figure 5(c), this results in an increase by order of magnitude at constant |r|, making the scheme comparable or better than the most sensitive available schemes [63-67].

However, the displayed increase in sensitivity requires resolving the incidence angle in the order of 0.01 deg. For the DBR structure displayed in Figure 5(c), the sensitivity drops by an order of magnitude with angle changing by just 0.02 deg, Figure 6(a). Although the presented system has superior sensitivity, it suffers from the same drawback as [66], as sensitivity is exceptionally dependent on the chosen incidence angle. This may result in unreliable sensitivity values, as most instruments cannot resolve at that level of precision.

To alleviate this stringent requirement, we exploit the phenomenon of merging charges introduced above. Interestingly, we show that by bringing two neighboring charges in close proximity, the reflection coefficient between them displays almost constant magnitude while keeping the strong phase jump, Figure 6(b)-(d). Furthermore, by fine-tuning the SiC thickness, we show that |r| can be steadily controlled and kept constant over an extended angular range, thus eliminating the stringent requirement on incidence angle and providing stable sensitivity. However, this comes at the cost of lower absolute sensitivity, because bringing two charges together requires lowering the overall Q-factor. A trade-off emerges between sensitivity of the system and robustness to incidence angle variations. For example, the structure in Figure 6(a) has 3 times larger absolute sensitivity than the structure in Fig. 6(d) (calculated for the same value of |r|). However, in most practical scenarios, it would be much easier to access the required angles of the latter structure and get the predicted sensitivity. Nevertheless, these results demonstrate that the topological features of EEs offer a versatile tool for sensing and can address different challenges of advanced sensing systems.

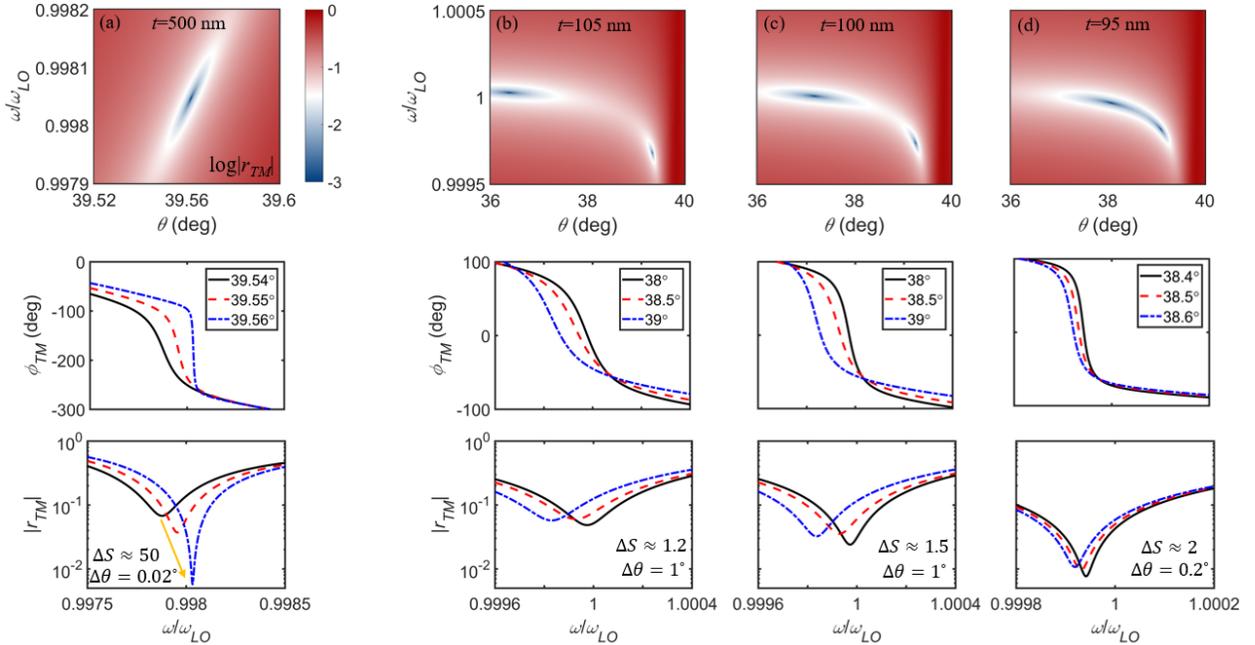

Figure 6. Sensitivity of the reflection coefficient and sensitivity for different incidence angles. (a) The reflection coefficient of the DBR structure in Figure 5(c) is extremely sensitive to small changes in incidence angle, resulting in one order of magnitude change in sensitivity for extremely small angle changes. (b)-(d) Near annihilation point sensing with stabilized sensitivity. The amplitude of the reflection coefficient is more robust to incidence angle variations, with the sensitivity not changing substantially for 1 deg changes in angle for (b) and (c), and 0.2 deg for (d).

## 3. Discussion

The theoretical results discussed in this paper extend the notion of topological scattering effects to planar structures and show that topological quantities can find their use in various scenarios of interest. Although the presented theory builds upon the topological features of EE-related phenomena, it is worth stressing the differences between the homogeneous planar systems analyzed here and EEs emerging due to the periodicity in photonic crystals (PCs). EEs in PCs come as isolated points in *k*-space, whereas in the homogeneous planar systems studied here, EEs generally emerge in rings in *k*-space and isolated points in real space. Thus, a topological charge cannot be defined by integrating in *k*-space like in PCs, but in real space (either in $\omega$-$d$ or $\omega$-$\theta$ space). Somewhat analogous to breaking spatial symmetries in PCs, which splits the EE into two half-integer charges [30], dissipation in our system lifts the degeneracy of two singularities of opposite integer charge, destroying the EE. Since all real systems have losses, this degeneracy is always lifted, whereas EEs in PCs remain topologically protected as long as spatial symmetries are not broken. Although PCs have more degrees of freedom to control the topological properties of EE scattering, we have shown that planar structures offer a versatile tool for intensity, phase and polarization control based on the topological nature of the reflection coefficient. Moreover, our proposed structures do not require complicated lithography fabrication process, as in the case of periodic structures.

The scattering matrix approach used here has also been utilized in a different area of topological photonics - Weyl physics [69], where topologically non-trivial states have been shown to have a non-zero winding number of the phase of scattering matrix eigenvalues [70]. This approach is especially valuable because the scattering matrix features can be observed experimentally, differently from the standard Hamiltonian approach. Furthermore, recent studies show that accounting for dissipation turns Weyl points into rings [71], somewhat analogous to what happens to EEs in our structures. A comprehensive study addressing Weyl physics in structures supporting both symmetry-protected and accidental EEs, and its effects on scattering/reflection, is yet to be done. As discussed in the previous paragraph, breaking spatial symmetries in PCs leads to splitting of EEs into half-integer, CP states. Similarly, breaking time-reversal symmetry by applying a magnetic field in plasma was shown to split Dirac points into Weyl points associated with CP and helical states [72]. Thus, applying a symmetry-breaking mechanism to our system may provide rich topological phenomena and establish a deeper connection between different branches of topological photonics.

To conclude, we have introduced a topological perspective on scattering singularities in planar systems supporting EEs, connecting several research areas. We have unveiled that singularities of the scattering matrix necessarily emerge from EEs and carry topological charges observable as phase vortices. Charge conservation was demonstrated by proposing a method of charge annihilation, where tuning the underlying radiation losses provides control over the features and motion of charges. It was shown that perfect absorption and phase vortices associated with these charges enable extreme control of the intensity, phase and polarization of the reflected waves. Based on these concepts, we have proposed applications for versatile polarization control and switching, as well as for phase sensing schemes that may be implemented using silicon carbide.

**METHODS**

**Analytical simulations**. The analytical response of the analyzed structures was calculated using transmission line and transfer matrix methods. Structures analyzed in Figures 1 and 2 can be represented by an equivalent transmission line circuit, taking into account polarization and transverse wavenumber. The transmission line method for PEC-backed slab of ENZ material yields the reflection coefficient singularities dispersion relation:

$$\tanh(jk_{1z}t) = \pm \frac{Z_0}{Z_1},$$

calculated in the $\omega_r$-$\omega_i$-$k_{\parallel}$ parameter space. The same method for the spacer structure yields reflection zero dispersion relation:

$$\tanh(jk_{2z}d) = \frac{Z_1}{Z_2}\left(\frac{Z_1 \tanh(jk_{1z}t) - Z_0}{Z_0 \tanh(jk_{1z}t) - Z_1}\right),$$

calculated in the $\omega_r$-$\omega_i$-$\theta$ parameter space. The TM and TE wave impedances of medium $n$ are calculated as $Z_n^{TM} = \cos\theta \sqrt{\mu_0/\varepsilon_0\varepsilon_n}$ and $Z_n^{TE} = \frac{1}{\cos\theta}\sqrt{\mu_0/\varepsilon_0\varepsilon_n}$. More details on the derivation can be found in supplementary materials.

**ACKNOWLEDGEMENTS**

The work described in this paper is conducted within the project NOCTURNO, which receives funding from the European Union's Horizon 2020 research and innovation programme under Grant No. 777714, as well as the ANTARES project that has received funding from the European Union's Horizon 2020 research and innovation program GA 739570. The work was also partially supported by the Department of Defense, the Air Force Office of Scientific Research, the National Science Foundation and the Simons Foundation.

**AUTHOR CONTRIBUTION**

Z.S, A.K. and N.J. conceived the idea. Z.S. performed the analytical simulations and wrote the manuscript. A.K, N.J., and A.A contributed to the manuscript and supervised the project.

# Supplementary material: Topological scattering singularities and embedded eigenstates for polarization control and sensing applications

Zarko Sakotic[1*], Alex Krasnok[2], Andrea Alú[2], and Nikolina Jankovic[1]

[1]BioSense Institute-Research Institute for Information Technologies in Biosystems, University of Novi Sad, Dr Zorana Djindjica 1a, 21101, Novi Sad, Serbia

[2]Advanced Science Research Center, City University of New York, New York, NY 10031, USA

### 4. Reflection coefficient

Throughout the paper, the reflection coefficient has been calculated using a transmission line model with a short-circuit at the end, making it equivalent to a one-port circuit backed with PEC. For a single slab of ENZ, the simple model in Figure S1(a) provides a condition for pole/zero dispersion.

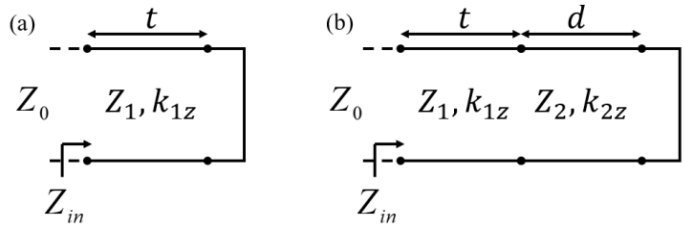

Figure S1: Transmission line model for a single slab and spacer structure.

$$Z_{in} = Z_1 \tanh(jk_{1z}t) \tag{s1}$$

$$r = \frac{Z_{in} - Z_0}{Z_{in} + Z_0} \tag{s2}$$

$$\tanh(jk_{1z}t) = \pm \frac{Z_0}{Z_1} \tag{s3}$$

For the spacer structure, following the same method, the reflection zero condition reads:

$$\tanh(jk_{2z}d) = \frac{Z_1}{Z_2}\left(\frac{Z_1 \tanh(jk_{1z}t) - Z_0}{Z_0 \tanh(jk_{1z}t) - Z_1}\right) \tag{s4}$$

For the realistic structures in the paper, the ABCD matrix method was used to obtain the reflection/transmission/absorption coefficients [73].

## 5. Charge annihilation

As demonstrated in the main text, charge conservation is confirmed through charge annihilation. We show here in more detail how charges behave and what role embedded eigenstates have. As charges emerge in pairs around the angle at which embedded eigenstate occurs, it is evident that these cannot annihilate each other (except for precisely EE with zero losses). Figure S2 shows how neighboring charges for the ENZ case annihilate in a system with several EEs. Starting from a system with $d=\lambda_0$ and $t=250$ nm, we reduce the top layer thickness $t$ and lower the Q-factor of the modes. This system supports 3 orders of the Fabry-Perot resonances and consequently supports 3 non-symmetry protected EEs.

Charges associated with symmetry-protected EE and the first accidental one at a small angle ($EE_1$) are of lower Q-factors, thus annihilate first. As thickness is further reduced, charges between $EE_1$ and $EE_2$, and finally, the pair between $EE_2$ and $EE_3$ annihilates. Thus the resonant angle of EE behaves as an impenetrable wall for these charges. The last unpaired charge of $EE_3$ remains in the spectrum and cannot be annihilated unless a 4$^{th}$ order EE is made available.

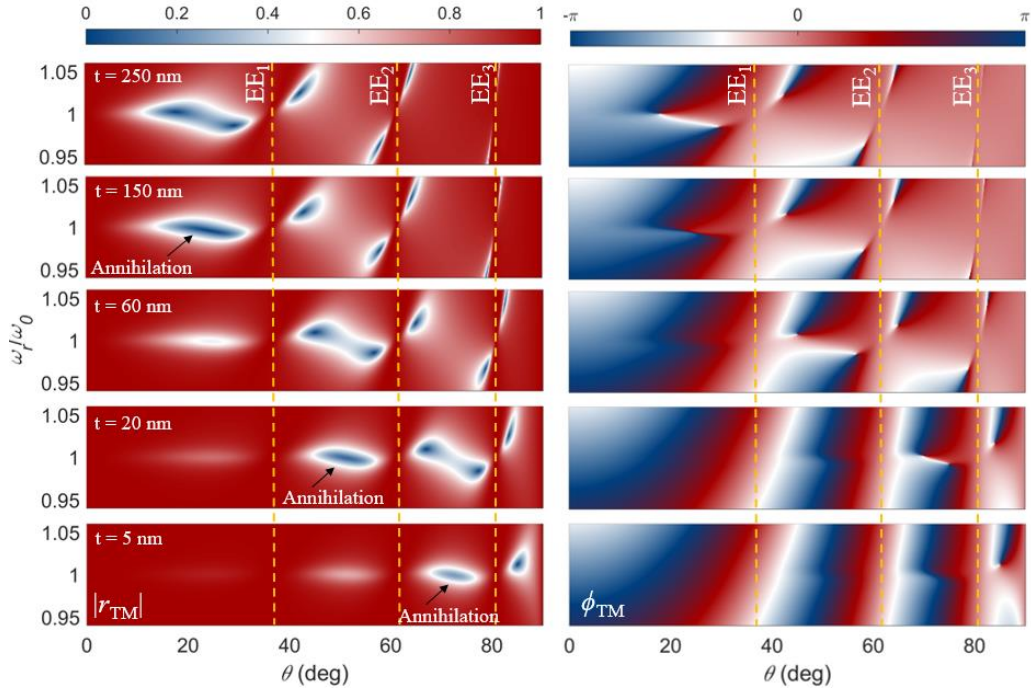

Figure S2: Annihilation of charges in a system with multiple embedded eigenstates.

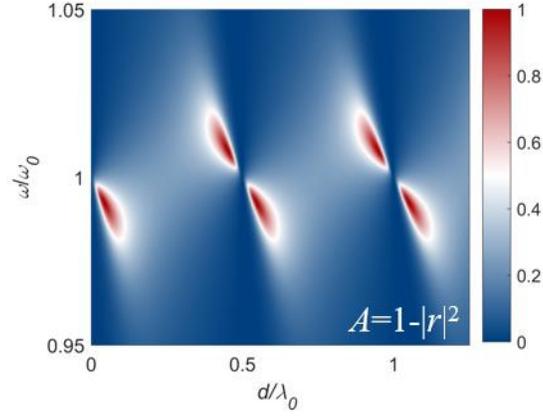

Figure S3. Perfect absorption for normally incidence waves around the ENP region.

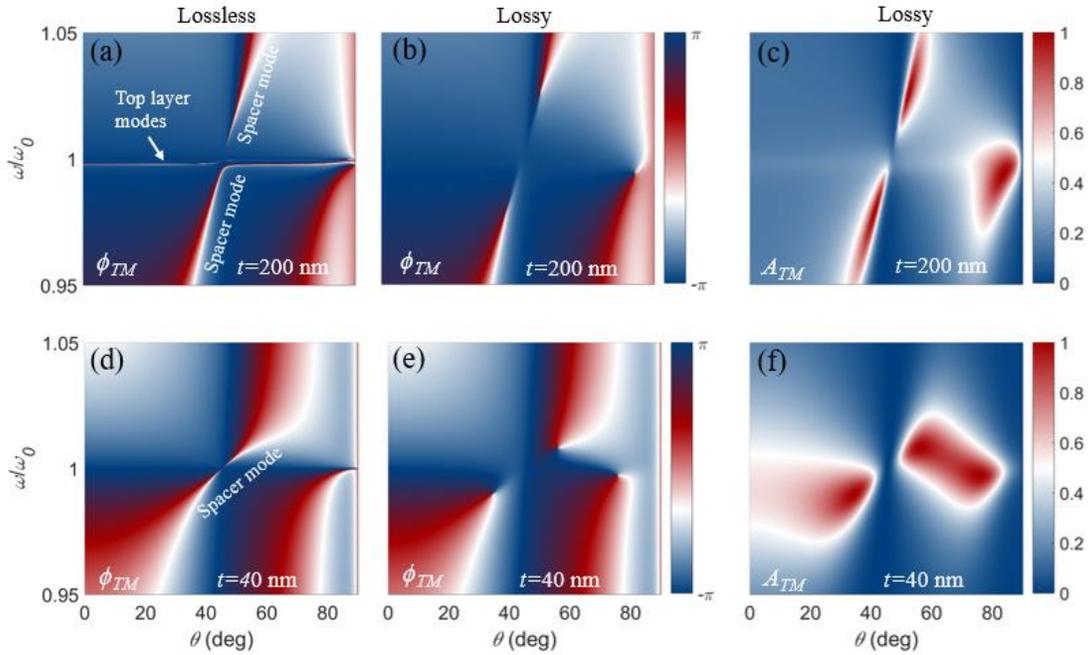

Figure S4. ENP scattering features. (a) For $t=200$ nm, propagating modes in the top layer are visible in the lossless case, along with the spacer mode. (b) Phase vortices appear in lossy case, top layer modes are strongly suppressed and not visible. (c) Perfect absorption points appear in the lossy case. (d)-(f) Same as (a)-(c) for $t=40$ nm.

## 6. ENP case scattering features

In terms of scattering/absorption properties, ENZ can offer various opportunities, however it cannot absorb normally incident waves due to the TM plasmonic nature of the underlying modes. On the other hand, ENP-based structures analyzed in Figure 2 of the main text do not have that drawback, and can absorb at normal incidence for different resonator thicknesses, Figure S3.

Another important aspect of ENP-based structures is that, at $\omega_0$, the permittivity changes from one extreme to another, drastically changing the nature of the top layer in the spacer structures considered throughout the main text. Namely, in the lossless case, by approaching the Lorentz resonance from the right $\omega \rightarrow \omega_0^+$, the permittivity of the top layer goes to $-\infty$, providing metallic, and ultimately PEC character to the layer and in this way enabling perfect light trapping of the Fabry-Perot mode existing in the spacer. However, by approaching the resonant frequency from the left side $\omega \rightarrow \omega_0^-$, permittivity of the top layer goes to $+\infty$, allowing for an infinite number of propagating solutions in the top layer which complicates the picture in terms of scattering, Figure S4 (a). By reducing the thickness of the top layer, these modes are confined very close to the resonant frequency $\omega_0$, and thus are not visible in the graphs in Figure S4 (d) and Figure 2 (i,l) in the main text. However, as shown in Figure S4 (b,c,e,f), applying losses considered in main text leads to strong suppression of the modes propagating in the top layer, both for thicker and thiner top layers. The emergence of three charges remains apparent in both cases, and there is little difference in terms singularities associated with the spacer mode. Nevertheless, certain material permittivies and thicknesses might yield a more complicated scattering picture around the resonance frequency, and require a more careful analysis of the topological scattering features.

### 7. Polarization definitions and output polarization state dependence on material loss

To assess the polarization properties of the structure, we define the transverse electric field in the plane normal to the incident *k*-vector, as sketched in figure S5 (a):

$$E^i = E_V + E_H = \begin{bmatrix} a \\ be^{j\delta_i} \end{bmatrix} e^{-j(k_{0x}x + k_{0z}z)}, \tag{s5}$$

where the incident electric field consists of vertically and horizontally polarized components with amplitudes *a* and *b* with the phase difference $\delta_i$. The compact way to describe the polarization is the Jones vector:

$$J^i = \begin{bmatrix} a \\ be^{j\delta_i} \end{bmatrix}. \tag{s6}$$

The reflected field now has

$$E^r = \begin{bmatrix} r_{TE} & r_{TE-TM} \\ r_{TM-TE} & r_{TM} \end{bmatrix} \begin{bmatrix} a \\ be^{j\delta_i} \end{bmatrix} e^{-j(k_{0x}x - k_{0z}z)}, \tag{s7}$$

where the reflection matrix describes the connection between the incident and the reflected field. As there is no cross-coupling mechanism between orthogonal polarizations, the terms $r_{TE-TM}$ and $r_{TM-TE}$ are equal to zero. The output Jones vector can then compactly be written as

$$J^o = \begin{bmatrix} r_{TE}a \\ r_{TM}be^{j\delta_i} \end{bmatrix} = \begin{bmatrix} |r_{TE}|a \\ |r_{TM}|e^{j\delta_r}be^{j\delta_i} \end{bmatrix} = \begin{bmatrix} A \\ Be^{j\delta_o} \end{bmatrix} \quad (s8)$$

with $A$, $B$, and $\delta_o$ being the output amplitudes and phase difference of orthogonal polarizations, and $\delta_r$ being the phase retardation between H and V-polarizations that the structure introduces.

As discussed in the main text, different topologically enabled polarization features emerge in the vicinity of the embedded eigenstate. Thus there is a possibility of external control and switching of polarization by changing the permittivity (loss) of the top layer, where small changes would significantly change the output polarization. To demonstrate the sensitivity of the output polarization to material losses and its potential for polarization switching implementation, we plot the output polarization in two spectral points for different values of the SiC loss parameter γ, Figure S6. We excite the system at the singular points of vertical (TE) and horizontal (TM) polarization at starting loss value $\gamma_1$=0.04 THz. By increasing loss to $\gamma_2$=0.08 THz and $\gamma_3$=0.012 THz, output polarization changes notably for the same frequency and angle of excitation, indicating switching capabilities.

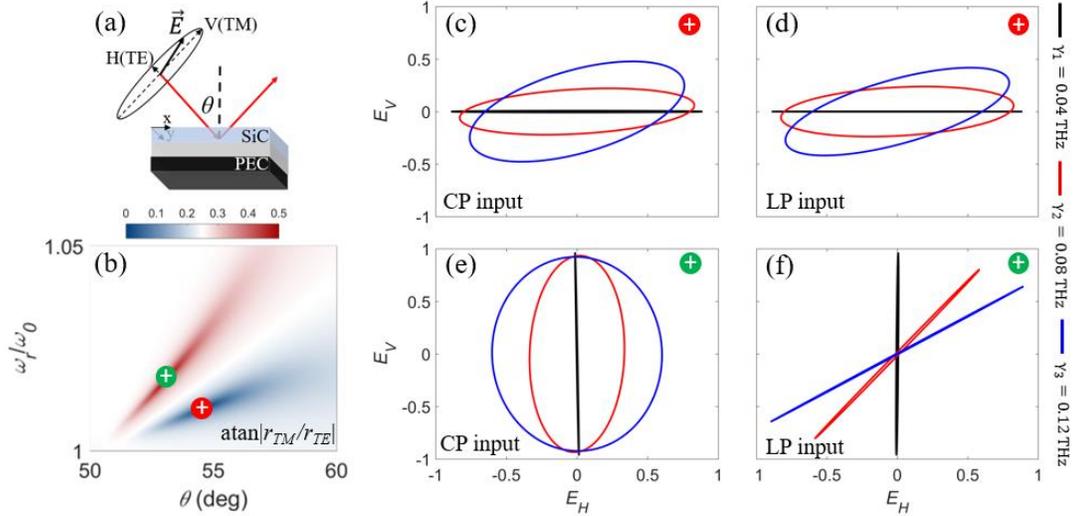

Figure S6. (a) Sketch of the polarization plane and the reflection problem under analysis. (b) Ellipsometric parameter $\psi = tan^{-1}(r_{TM}/r_{TE})$ where zeros of different polarizations are highlighted. (c) Output polarization dependence on the material loss parameter γ for CP input at TM zero. (d) 45 deg LP input at TM zero. (e) CP input at TE zero. (f) 45 deg LP input at TE zero.